\documentclass[aps,pra,twocolumn,superscriptaddress,amsmath,amssymb,showpacs,floatfix]{revtex4}
\usepackage{latexsym}
\usepackage{color}
\usepackage{graphicx}
\usepackage{bm}
\usepackage{times}
\renewcommand{\vec}[1]{\mathbf{#1}}
\newcommand{\ii}{\mathrm{i}}
\renewcommand{\d}{\mathrm{d}}
\newcommand{\tc}{T_{\cal C}}
\newcommand{\dc}{\delta_{\cal C}}
\newcommand{\intc}{\int_{\cal C}}

\newcommand{\lowerbossphantom}{\vphantom{\bar{{x}}}}
\newcommand{\upperbossphantom}{\vphantom{\dagger}}
\newcommand{\Aop}[1]{\ensuremath{c_{#1\lowerbossphantom}^{\upperbossphantom}}}
\newcommand{\Cop}[1]{\ensuremath{c_{#1\lowerbossphantom}^{\dagger\upperbossphantom}}}
\begin{document}
\title{Stopping dynamics of ions passing through correlated honeycomb clusters}
\author{Karsten Balzer}
\email[E-mail address: ]{balzer@rz.uni-kiel.de}
\affiliation{Rechenzentrum, Christian-Albrechts-Universit\"{a}t zu Kiel, Ludewig-Meyn-Strasse 4, 24118 Kiel, Germany}
\author{Niclas Schl\"{u}nzen}
\affiliation{Institut f\"{u}r Theoretische Physik und Astrophysik, Christian-Albrechts-Universit\"{a}t zu Kiel, Leibnizstrasse 15, 24098, Kiel, Germany}
\author{Michael Bonitz}
\affiliation{Institut f\"{u}r Theoretische Physik und Astrophysik, Christian-Albrechts-Universit\"{a}t zu Kiel, Leibnizstrasse 15, 24098, Kiel, Germany}
\begin{abstract}
A combined nonequilibrium Green functions--Ehrenfest dynamics approach is developed that
allows for a time-dependent study of the energy loss of a charged particle penetrating
a strongly correlated system at zero and finite temperature. Numerical results are presented
for finite inhomogeneous two-dimensional Fermi--Hubbard models, where the many-electron dynamics
in the target are treated fully quantum mechanically and the motion of the projectile is
treated classically. The simulations are based on the solution of the two-time Dyson
(Keldysh--Kadanoff--Baym) equations using the second-order Born, third-order and T-matrix
approximations of the self-energy. As application, we consider protons and helium nuclei
with a kinetic energy between~$1$ and $500$\,keV/u passing through planar fragments of the
two-dimensional honeycomb lattice and, in particular, examine the influence of
electron-electron correlations on the energy exchange between projectile and electron
system. We investigate the time dependence of the projectile's kinetic energy (stopping power),
the electron density, the double occupancy and the photoemission spectrum. Finally, we show
that, for a suitable choice of the Hubbard model parameters, the results for the stopping power
are in fair agreement with ab-initio simulations for particle irradiation of single-layer
graphene.
\end{abstract}
\maketitle

\section{\label{sec.intro}Introduction}

The interaction of particles with matter is a fundamental aspect of physics and allows
one to measure their properties in colliding-beam or beam-target experiments. Conversely,
the irradiation of matter by particles can be used also as a diagnostic tool to probe the
static and dynamic properties of matter itself. In soft collisions of heavy charged particles,
such as ions, with a solid, typically the electrostatic force, i.e., the Coulomb interaction,
has the largest impact, leading to excitation and ionization of electrons in the target material
and thus to the loss of kinetic energy of the projectile~\cite{sigmund06}. For nonrelativistic
projectile velocities of the order of or larger than the Fermi velocity ($\sim10^6$\,m/s
in metals), theoretical approaches based on scattering theory~\cite{nagy98} or on the response
functions of the homogeneous electron gas~\cite{pitarke95}, can give a quantitative description
of the energy transferred during the collision process but neglect the precise atomic composition
of the target.

Regarding first-principles modeling in the same velocity regime, recent theoretical progress
is due to time-dependent density functional theory (TDDFT), which has been applied to describe
the slowing down of charged particles in a variety of solids, including metals~\cite{quijada07,zeb12,schleife15},
semimetals~\cite{ojanperae14,zhao15} and clusters~\cite{bubin12,mao14}, narrow-band-gap
semiconductors~\cite{ullah15} and insulators~\cite{pruneda07,zeb13}. Taking into account
primarily the excitation of valence electrons, these simulations yield satisfactory results
for the electronic stopping power (the transfer of energy to the electronic degrees of freedom
per unit length traveled by the projectile) and work for a wide range of impact energies. On
the other hand, one can quite generally determine the stopping power of energetic ions in matter
using the SRIM code~\cite{srim08}, which uses the binary collision approximation in combination
with an averaging over a large range of experimental situations to provide energy loss and range
tables for many materials and gaseous targets. In principle, TDDFT and SRIM can include effects
of electron-electron correlations on the stopping behavior, either by using exchange-correlation
potentials beyond the local density approximation in TDDFT, e.g., \cite{nazarov05,nazarov07}, or
by including static exchange and correlation contributions to the interaction energy between
overlapping electron shells in SRIM. Despite these capabilities, both methods have, however,
difficulties to approach \emph{strong} Coulomb correlations, which are crucial, e.g., in transition
metal oxides~\cite{anisimov97} or specific organic materials~\cite{singla2015}. In addition, we
note that SRIM does account neither for the exact crystal structure of the material nor dynamic
(i.e., time-dependent) changes in the target during the collision process, which limits its
applicability.

It is, therefore, interesting to consider an alternative approach to the stopping power that does
not have these limitations: nonequilibrium Green functions~(NEGF)~\cite{kadanoff62,stefanucci13.cup}.
This method allows one to systematically include electron-electron correlations via a time-dependent
many-body self-energy, and it has recently successfully been applied to strongly correlated lattice
systems as well~\cite{schluenzen16}. Particular advantages of the NEGF approach are that it is not
limited to either weak or strong coupling and that it is particularly well suited to study finite-sized
clusters and spatially inhomogeneous systems on a self-consistent footing.  While the NEGF approach
is computationally very demanding, in recent years efficient numerical schemes have been developed
to solve the underlying Keldysh--Kadanoff--Baym equations (KBE)~\cite{dahlen07,stan09,balzer10.pra1,balzer10.pra2,garny10,balzer13.lnp,latini14,hermanns14}. 

Here, we extend the NEGF approach by including the interaction with a classical projectile using
an Ehrenfest-type approach that is well established in TDDFT simulations. Our goal is to develop a
full time-dependent and space-resolved description, which is necessary as the projectile induces
local time-dependent changes to the electron density and to the local band structure. This allows
us, in particular, to consider finite clusters of size $L$, which are of substantial current interest.
Furthermore, we study the size dependence of the response to the projectile. At the same time,
the thermodynamic limit of the stopping power (cluster size $L$ approaching infinity) is more
difficult and expansive to obtain, as it requires an extrapolation of results for different $L$.
Nevertheless, we obtain good agreement with existing results for macroscopic systems.

To implement this approach, we choose, as a first application, the energy deposition of simple ions
(protons and alpha particles), where there is no inter-atomic electron dynamics, in planar two-dimensional
honeycomb clusters, in which the electron dynamics are well described in terms of a Fermi--Hubbard
model. To investigate the importance of electronic correlation effects, we vary the coupling strength
from small to moderate values (up to $U/J=4$) and test various self-energy approximations, such as
the second Born and the much more involved T-matrix approximation. The results are compared to
mean-field (Hartree) results, which are provided by the same NEGF program.

The paper is organized as follows. In Sec.~\ref{sec.comp.setup}, we define the model Hamiltonian, discuss
the interaction potential between projectile and target, and describe the self-consistent computational
scheme, which allows us to calculate the correlated electron dynamics on the honeycomb clusters. In
Sec.~\ref{sec.lattice.prop}, we review the equilibrium properties of the target system, which are sensitive
to correlations, and then present the main results for the stopping dynamics in Sec.~\ref{sec.stopping.dyn}.
Here, we primarily focus on the effect of electron-electron correlations on the energy transfer,
analyze the time-dependent collision process for a wide range of projectile velocities and consider
different initial states and temperatures. In Sec.~\ref{sec.graphene}, we finally discuss the
application of the used model to graphene and conclude the paper with Sec.~\ref{sec.conclusions},
outlining possible future work.


\section{\label{sec.comp.setup}Computational setup}

\subsection{\label{subsec.model}Model}

To study the stopping dynamics of a classical charged particle which passes through a
(strongly) correlated system, we consider a finite lattice of electrons described by a
single-band Fermi--Hubbard model and monitor the transfer of energy during the collision
process. Taken as a whole, the lattice system is electrically neutral, i.e., the electronic
charges are compensated by corresponding opposite charges located at the site coordinates
$\vec{R}_i$. The general stopping mechanism is mediated by the bare Coulomb interaction
between the projectile, the fixed background charges and the target electrons which are
initially in equilibrium. Throughout, we focus on positively charged ions as projectiles,
which,  when approaching the lattice, induce a confinement potential to the electrons and
thus initiate a nonequilibrium electron dynamics. In turn, the ions (of mass $m_\textup{p}$
and charge $Z_\textup{p}e$) react to any charge redistribution on the lattice and change
their position and kinetic energy accordingly.

\begin{figure}[t]
 \includegraphics[width=0.483\textwidth]{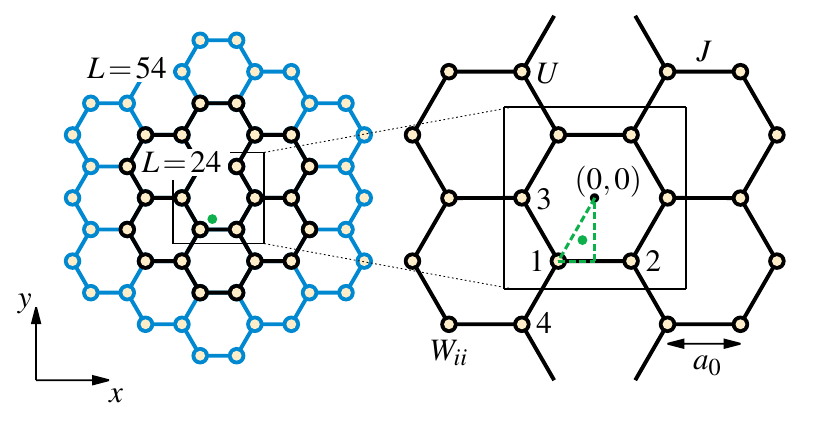}
 \caption{(Color online) Lattice structure of circular honeycomb clusters with $L=24$
 (black) and $54$ (blue) sites. The green point indicates the position where the projectile
 hits the lattice plain. For further reference, we label four sites in the center of the
 clusters, where we will monitor the time-dependent electron density in~Sec.~\ref{subsec.td.density}.
 Furthermore, $a_0$ denotes the lattice spacing, $J$ ($U$) is the nearest-neighbor hopping
 (the on-site interaction), and $W_{ii}$ is the local energy defined in Eq.~(\ref{eq.ham2}).}
 \label{fig.lattice}
\end{figure}

As lattice systems, we choose circular honeycomb clusters, which are oriented in the
$xy$-plane and have a finite number of honeycombs, yielding in total $L$ sites, see
Fig.~\ref{fig.lattice} for an illustration. We consider a half-filled system in the
paramagnetic phase and, to generate realistic results, set the lattice spacing to $a_0=1.42$\,{\AA},
which corresponds to the carbon-carbon bond length in graphene~\cite{katsnelson12}.
Using a nearest neighbor-hopping $J$ and an on-site Coulomb repulsion $U$, the Hamiltonian
for the lattice electrons is then given by
\begin{align}
\label{eq.ham1}
 H_\textup{e}(t)&=-J\sum_{\langle i,j\rangle,\sigma} \Cop{i\sigma} \Aop{j\sigma}+U\sum_{i}(n_{i\uparrow}-\tfrac{1}{2})(n_{i\downarrow}-\tfrac{1}{2})&\nonumber\\
&\hspace{1pc}+\sum_{ij,\sigma}W_{ij}(t) \Cop{i\sigma} \Aop{j\sigma}\,,
\end{align}
where the operator $\Cop{i\sigma}$ ($\Aop{i\sigma}$) creates (annihilates) an electron
with spin $\sigma$ on site $i$, $n_{i\sigma}=\Cop{i\sigma} \Aop{i\sigma}$ denotes the
electron density, and $W_{ij}$ are the matrix elements of the confinement potential
induced by the projectile. In Sec.~\ref{sec.stopping.dyn}, we imply localized electronic 
ave functions $\varphi_i(\vec{r})\propto\delta(\vec{r}-\vec{R}_i)$, for which we can resort
to the diagonal components of this potential:
\begin{align}
\label{eq.ham2}
 W_{ii}(t)=-\frac{e^2}{4\pi\epsilon_0}\frac{Z_\textup{p}}{|\vec{r}_\textup{p}(t)-\vec{R}_i|}\,,
\end{align}
where $\vec{r}_\textup{p}(t)$ denotes the time-dependent position of the projectile, $-e$ is
the electron charge and $\epsilon_0$ the vacuum permittivity. Moreover, in Sec.~\ref{sec.graphene},
we improve this model by including also terms $W_{ij}(t)$ with $|i-j|=1$, which locally renormalize
the nearest-neighbor hopping, cf.~Eq.~(\ref{eq.hoppingrenormalization}).

For convenience, we measure $J$ and $U$ in electron volts, define $U/J$ as the
interaction strength for the electrons and use $t_0=\hbar/J$ as the unit of time.
Unless otherwise stated, we use $J=2.8$\,eV (which is typical for graphene~\cite{schueler13})
to fix the time scale.

\subsection{\label{subsec.comp.method}Computational Method}

To compute the classical motion of the projectile with an initial velocity
$\d \vec{r}_\textup{p}/\d t=(0,0,v_z)$, we solve Newton's equation with the total
 potential
\begin{align}
\label{eq.ham3}
V(\vec{r}_\textup{p},t)=\frac{e^2}{4\pi\epsilon_0}\sum_{i}\frac{Z_\textup{p}Z_i(t)}{|\vec{r}_\textup{p}(t)-\vec{R}_i|}
\end{align}
created by all lattice charges (here, $Z_i(t)=1-\sum_\sigma \langle n_{i\sigma}\rangle(t)$
denotes the net charge on the lattice site $i$). For the solution we use a three-dimensional
velocity-Verlet algorithm. Motivated by TDDFT calculations~\cite{zhao15}, we set the initial
position of the incident ion to $\vec{r}_{\textup{p}}=\left(-\tfrac{1}{6}a_0,-\tfrac{\sqrt{3}}{3}a_0,-z\right)$,
see the centroid point of the green dashed triangle in Fig.~\ref{fig.lattice}. These coordinates
have been found to give similar stopping results for the highly symmetric honeycomb lattice
compared to calculations, where one averages over many different collision sites. This allows
us to avoid averaging over many trajectories and to directly compare to previous TDDFT results (Sec.~\ref{sec.graphene}).
Furthermore, the initial $z$-position is chosen such that the measured energy transfer becomes
independent of the initial conditions (typically $z\gtrsim10a_0$).

To compute the correlated time evolution of the lattice electrons, we use a
quantum statistical approach based on the one-particle nonequilibrium Green
function (NEGF)
\begin{align}
\label{eq.negf}
 G_{ij\sigma}(t,t')=-\frac{\ii}{\hbar}\langle \tc \Aop{i\sigma}(t) \Cop{j\sigma}(t')\rangle\,,
\end{align}
which is defined on the Keldysh time contour $\cal C$~\cite{keldysh64} and can be
interpreted as a two-time generalization of the one-particle density matrix,
\begin{align}
 \langle\rho_{ij\sigma}\rangle(t)=\langle \Cop{i\sigma}\Aop{j\sigma}\rangle=-\ii\hbar\,G_{ji\sigma}(t,t^+)\,.
\end{align}
On the contour, $\tc$ furthermore denotes the time-ordering operator,
$\langle\tc\ldots\rangle=\textup{tr}[\tc \exp(S)\ldots]/\textup{tr}[\tc\exp(S)]$
with ${S=-\ii/\hbar \intc \d s\,H_\textup{e}(s)}$ defines the ensemble average, and
the notation $t^+$ means, that the time $t^+$ is infinitesimally larger along $\cal C$
than $t$. The equations of motion of the greater and less components of the NEGF (\ref{eq.negf}),
\begin{align}
 G^>_{ij\sigma}(t,t')&=-\frac{\ii}{\hbar}\langle \Aop{i\sigma}(t)\Cop{j\sigma}(t')\rangle\,,\\
 G^<_{ij\sigma}(t,t')&=\frac{\ii}{\hbar}\langle \Cop{j\sigma}(t')\Aop{i\sigma}(t)\rangle\,,\nonumber
\end{align}
follow from the time evolution of the creation and annihilation operators in the
Heisenberg representation and are known as the two-time Keldysh--Kadanoff--Baym
equation (KBE)~\cite{kadanoff62,stan09,stefanucci13.cup}:
\begin{align}
\label{eq.kbe}
 \sum_k&[\ii\hbar\,\partial_t\delta_{ik}-h_{ik\sigma}(t)]G_{kj\sigma}^\gtrless(t,t')&\\
 &=\dc(t,t')\delta_{ij}+\sum_{k}\left\{\intc \d s\,\Sigma_{ik\sigma}(t,s)G_{kj\sigma}(s,t')\right\}^\gtrless\,.\nonumber
\end{align}
Here, $\dc$ denotes the delta function on the contour, and $h_{ij\sigma}(t)$ is the
time-dependent effective one-particle Hamiltonian, which explicitly includes the
Hartree contribution to the electron-electron interaction, i.e.,
\begin{align}
\label{eq.hij}
h_{ij\sigma}(t)=-\underbrace{J\delta_{\langle i,j\rangle}}_{=J_{ij}}+[W_{ii}(t)+U(\langle n_{i\bar{\sigma}}\rangle(t)-\tfrac{1}{2})]\delta_{ij}\,,
\end{align}
with the density $\langle n_{i\sigma}\rangle(t)=-\ii \hbar G_{ii\sigma}^<(t,t)$.
On the right-hand side of Eq.~(\ref{eq.kbe}), the contour integral defines the memory
kernel of the KBE, in which $\Sigma_{ij\sigma}(t,t')$ denotes the correlation part
of the self-energy [i.e., the mean-field part is excluded as it is contained in
Eq.~(\ref{eq.hij})]. Systematic expressions for the self-energy can be constructed
by many-body perturbation theory, e.g.,  using diagram techniques~\cite{stefanucci13.cup,schluenzen16.cpp}.
Below, we treat the correlation self-energy $\Sigma$ in different approximations,
which conserve particle number, momentum and energy.

\subsection{\label{subsec.mb.approx}Many-body approximations}

We consider the correlation self-energy $\Sigma$ in the following approximations:
\begin{enumerate}
\item 
As the simplest self-energy beyond the  (Hartree) mean-field level,
we consider the second-order Born approximation (2B),
 \begin{align}
 \label{eq.sigma.2b}
  \Sigma_{ij\sigma}^{\textup{2B},\lessgtr}(t,t')=\hbar^2U^2G^\lessgtr_{ji\sigma}(t,t') G^\lessgtr_{ji\bar{\sigma}}(t,t') G^\gtrless_{ij\bar{\sigma}}(t',t)\,,
 \end{align}
which includes all irreducible diagrams of second order in the interaction $U$.
Aside from the full evaluation of this self-energy, we will consider, in addition,
the \emph{local} (in space) second Born approximation, which includes only the
diagonal components $\Sigma^{\textup{2B}}_{ii\sigma}$ of the self-energy~(\ref{eq.sigma.2b}).
This approximation substantially reduces the numerical complexity, as it allows to
solve the KBE via particularly efficient schemes~\cite{balzer14.aux,balzer16.aux,gramsch15}.
We note that the 2B approximation is a perturbation theory result and, therefore,
becomes less accurate when $U$ increases.
\item
We consider the particle-particle T-matrix (TM) self-energy, which sums up the whole
Born series including diagrams of all orders in $U$ and is given by~\cite{puigvonfriesen09}
 \begin{align}
 \label{eq.sigma.tm}
  \Sigma_{ij\sigma}^{\textup{TM}}(t,t')=\ii\hbar\,T_{ij}(t,t')G_{ji\bar{\sigma}}(t',t)\,,
 \end{align}
with the effective interaction
 \begin{align}
  T_{ij}(t,t')=&-\ii\hbar\,U^2 G_{ij\sigma}(t,t')G_{ij\bar{\sigma}}(t,t')\nonumber\\
  &+\ii\hbar\,U\sum_k\intc \d s\,G_{ik\sigma}(t,s)G_{ik\bar{\sigma}}(t,s)T_{kj}(s,t')\,.\nonumber
 \end{align}
The T-matrix approximation has been found to perform very well in the regime of small
(or large) density, i.e., away from half-filling~\cite{puigvonfriesen09,schluenzen16.cpp,hermanns16}.
If the number of electrons and holes become comparable, however, particle-hole interaction
processes gain in importance, which are not captured by the particle-particle T-matrix. 
\item
In order to accurately treat strongly correlated systems, we also consider the third-order
approximation~\cite{hermanns16}, which exactly takes into account all self-energy
contributions up to $\mathcal{O}\left(U^3\right)$. This approximation has been found advantageous around
half-filling, in particular, for small to moderate interaction strengths~\cite{hermanns16}.
\item We also consider the generalized Kadanoff--Baym ansatz (GKBA) of Lipavsk\'{y} et al.~\cite{lipavsky86},
which has recently attracted growing attention~\cite{latini14,hermanns14,perfetto15,schluenzen16.cpp}.
It provides a way to reduce the numerical effort of the computation of the NEGF, while
still accurately accounting for particle number and energy conservation and correlations.
At the same time, it substantially reduces the computational effort, because the two-time
Green function $G_{ij\sigma}(t,t')$ is reconstructed from its time-diagonal value. Here,
we will apply the GKBA to the second-order Born self-energy using mean-field type propagators
(HF-GKBA), for details see Refs.~\cite{hermanns14,schluenzen16.cpp}. This allows us to
increase the simulation duration and extend the calculations to lower projectile energies,
see Sec.~\ref{subsec.gkba}.
\end{enumerate}

With these self-energies, the KBE~(\ref{eq.kbe}) is solved together with its adjoint
equation by a self-consistent time propagation scheme in the two-time plane, starting
from a given initial state Green function at $t,t'=0$. For details on the numerical
solution of the two-time KBE including the above approximations, we refer the reader to
Refs.~\cite{stan09,balzer10.pra2,balzer13.lnp,schluenzen16.cpp}. To investigate the
influence of the initial state of the many-electron system on the dynamics and the
energy loss of the projectile, we consider two relevant cases. For the example of the
system being initially in the ground state [we set $k_\textup{B}T=\beta^{-1}=0.01$\,eV
and note that the case of finite temperature is discussed separately in Sec.~\ref{subsec.finite.temp}],
we consider:
\begin{itemize}
\item [(A)] the stationary correlated equilibrium state, which is formed via a relaxation
that starts in the Hartree ground state (see Sec.~\ref{sec.lattice.prop} for details). This
is, of course, only an approximation to the true ground state but it substantially reduces
the computation time.
\item [(B)] the fully correlated ground state. It is obtained via a time-dependent procedure
(adiabatic switch-on of the interaction $U$), see, e.g., Refs.~\cite{rios11, schluenzen16.cpp}
for details.
\end{itemize}
For the case of half-filling (chemical potential $\mu=0$), the Hartree ground state of
Hamiltonian (\ref{eq.ham1}) [with $W_{ij}\equiv0$] is independent of $U$ and is given by
the density matrix
\begin{align}
 \langle\rho_{ij\sigma}\rangle(t=0)&=-\ii\hbar\,G^<_{ji\sigma}(0,0)&\nonumber\\
 \label{eq.hartree}
 &=-\ii\hbar\sum_k v_{ki}^*v_{kj}f_{\beta}(\epsilon_k,\mu)\,,
\end{align}
where $\epsilon_k$ ($\vec{v}_{k}$) are  the eigenvalues (eigenvectors) of the
hopping matrix $(\vec{J})_{ij}=J\delta_{\langle i,j\rangle}$, and $f_\beta(\epsilon,\mu)=1/(\textup{e}^{\beta(\epsilon-\mu)}+1)$
is the Fermi--Dirac distribution.


\section{\label{sec.lattice.prop}Lattice properties prior to the impact}

In this section, we solve the KBE~(\ref{eq.kbe}) without the incident projectile and
compute central equilibrium properties of the honeycomb clusters primarily in the local
second Born approximation. First, we analyze the double occupations,
\begin{align}
\langle d_i\rangle=\langle n_{i\uparrow} n_{i\downarrow}\rangle=-\frac{\ii\hbar}{ U}\sum_k\intc \d s\,\Sigma_{ik\sigma}(t,s) G_{ki\sigma}(s,t^+)\,,
\end{align}
on the lattice sites $i$, which contain important information about the correlations in
the system.

\begin{figure}[b]
\includegraphics[width=0.483\textwidth]{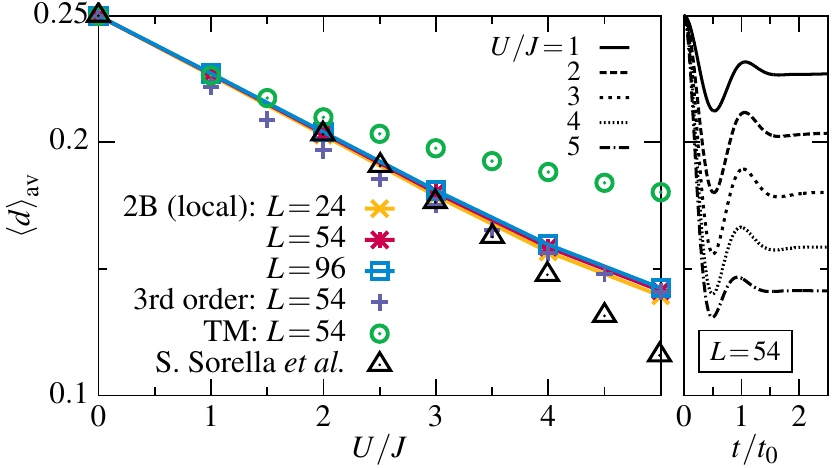}
 \caption{(Color online) Left panel:~Average double occupation ${\langle d\rangle}_\textup{av}$ on the
 honeycomb clusters with $L=24$, $54$ and $96$ sites for different interaction strengths $U/J$
 in the local second Born, third-order and T-matrix approximations, using the initial state~(A).
 The black triangles correspond to exact data for the extended honeycomb lattice (taken from
 Ref.~\cite{sorella92}). Right panel:~time evolution of ${\langle d\rangle}_\textup{av}$ for the local
 second Born calculations starting from the Hartree ground state [Eq.~(\ref{eq.hartree})],
 for which ${\langle d\rangle}_\textup{av}=\langle d_i\rangle=\langle n_{i\uparrow}\rangle\langle n_{i\downarrow}\rangle=0.25$.}
 \label{fig.equil.doubleocc}
\end{figure}

In Fig.~\ref{fig.equil.doubleocc}, we show results for the average double occupation
${\langle d\rangle}_\textup{av}=\tfrac{1}{L}\sum_i\langle n_{i\uparrow}n_{i\downarrow}\rangle$
on clusters of different size $L$, which is established over time, when the system is prepared
in the Hartree ground state [Eq.~(\ref{eq.hartree}), case (A)]. We find that the emerging double
occupation is practically independent of the system size, which is even the case for larger values
of the interaction strength. The value ${\langle d\rangle}_\textup{av}$ is, up to $U/J\lesssim4$,
in reasonable agreement with exact quantum Monte-Carlo data (black triangles) for the extended
honeycomb lattice~\cite{sorella92}. The right panel of Fig.~\ref{fig.equil.doubleocc} gives details
on the time dependence of the double occupation during this (fictitious) relaxation. Clearly,
the sudden switch-on of the correlation part of the self-energy at $t=0$ leads to an oscillatory
transient response, after which the double occupation rapidly reaches a new (correlated) stationary
value. The site densities ($\langle n_{i\sigma}\rangle=0.5$) remain constant during this relaxation
because, we consider an undoped system with particle-hole symmetry. We note that this final state
is a stationary correlated state, which slightly differs from the correlated ground state as it
has a slightly larger total energy (due to correlation induced heating~\cite{bonitz96,semkat99}),
cf.~the spectral weight discussed below. Nevertheless, the excellent agreement with the reference
data confirms the reliability of this procedure, which is computationally efficient as it requires
comparatively few time steps. 

Finally, the stationary values of the double occupations allow us to test the accuracy of
the different approximations for the self-energy. The T-matrix result is accurate up to
about $U/J=1.5$ but for larger coupling starts to deviate from the reference. The third-order
approximation and the local second Born result are very close to each other and work substantially
better up to $U/J=3.5$. Since the correlated states (A) and (B) are particle-hole symmetric, in
an exact calculation, the third-order contributions to the self-energy would perfectly
cancel each other~\cite{schluenzen16,hermanns16,gebhard03}. Therefore, in the T-Matrix, the
leading term beyond second order becomes unbalanced, which explains the poor performance in
Fig.~\ref{fig.equil.doubleocc}. For the same reason, both, a full second Born calculation and
a third-order simulation would be exact up to $\mathcal{O}\left(U^3\right)$, which also is the
origin for the high accuracy of the local second Born results. These findings give us confidence
to use the comparatively simple local second Born approximation for most simulations below~\cite{phsymmetry}.

\begin{figure}[t]
\includegraphics[width=0.483\textwidth]{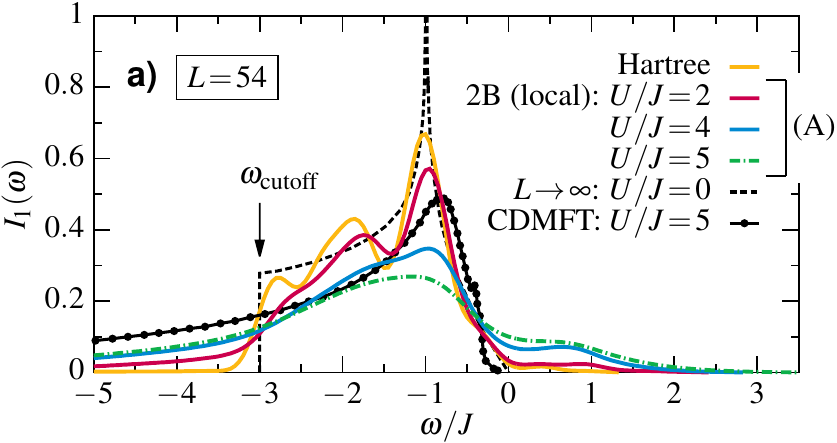}
\includegraphics[width=0.483\textwidth]{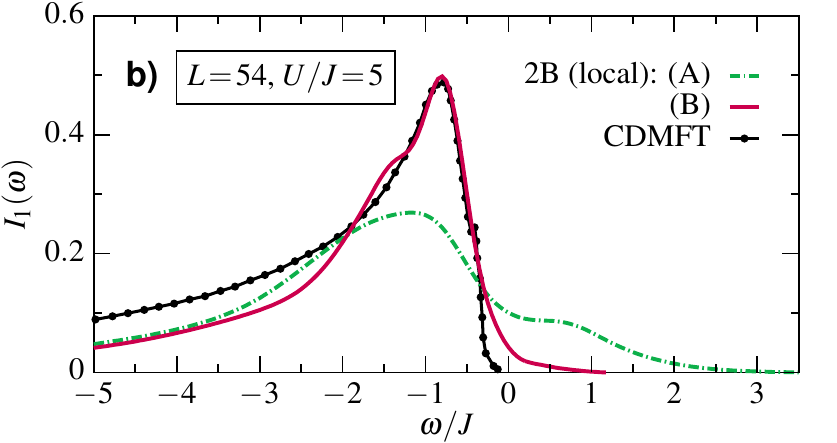}
 \caption{(Color online) a)~Photoemission signal $I_1(\omega)$ for the honeycomb cluster with
 $L=54$ sites in Hartree approximation ($U$-independent) and in local second Born approximation
 for different $U/J$, using the initial state (A) of Sec.~\ref{subsec.mb.approx}. Black dashed line:
 spectrum of the extended lattice at $U/J=0$ (e.g.~Refs.~\cite{neto09,he12}), black solid line:
 spectrum for $U/J=5$, as obtained from a cluster-DMFT calculation (from Ref.~\cite{liebsch11}
 and also shown in panel b)). b)~Comparison of the photoemission signal $I_1(\omega)$ at $U/J=5$
 for local second Born calculations with different initial states. Green line: spectrum (as in
 panel a)) for the initial state (A). Red line: correlated ground state (B).}
 \label{fig.equil.pes1}
\end{figure}

Second, we study the photoemission spectrum, which is directly obtained from the less
component $G_{ii\sigma}^<(t,t')$ of the nonequilibrium Green function, for details, see
the Appendix. In Fig.~\ref{fig.equil.pes1}, we present the photoemission signal $I_1(\omega)$
of the cluster with $L=54$ sites recorded at the central site $1$ [as labeled in Fig.~\ref{fig.lattice}].
We show results for different values of $U/J$ with a probe pulse arriving at some time
after the transient regime (for the computational details and the specific probe pulse
parameters, see Appendix~\ref{sec.appendix.spectrum}). In the case of half-filling, the Hartree
approximation ($\Sigma=0$) yields a photoemission signal with a few pronounced peaks which
are independent of $U/J$. On the other hand, correlations lead to an essential broadening
of the whole spectrum and, in particular, to single-particle energies beyond the cutoff
energy $\omega_{\textup{cutoff}}=-3J$ of the non-interacting system, see the black dashed
line and compare also with Fig.~\ref{fig.appendix.pes1} in Appendix~\ref{sec.appendix.spectrum}.
Moreover, this broadening is accompanied by a shift of the main peak (around $\omega=-J$)
towards the Fermi energy, $\omega_\textup{F}=0$, as a function of the interaction strength.
For $U/J=5$ (green dash-dotted line), we finally observe that the photoemission spectrum
becomes rather flat due to enhanced occupations of single-particle energies above the Fermi
level. These occupations originate from the fact that the equilibrated state is not the
ground state of the system. On the contrary, if we first prepare the  correlated ground
state~(B) [recall Sec.~\ref{subsec.mb.approx}] and then propagate the nonequilibrium Green
functions in time, we obtain a photoemission spectrum as shown by the red solid line in
Fig.~\ref{fig.equil.pes1}b. If we compare it, for example, to a cluster-DMFT (CDMFT) study~\cite{liebsch11}
for the extended honeycomb lattice (black solid line), we find a very good agreement.
However, the finite spectral resolution introduced by the probe pulse does not allow us
to recover the emergence of a small energy gap~\cite{he12}, which exists at finite on-site
interactions $U/J$.

In summary, we conclude from Figs.~\ref{fig.equil.doubleocc} and~\ref{fig.equil.pes1} that
already the local second Born approximation is able to capture important electron correlation
properties of the honeycomb clusters. As the considered equilibrium properties are adequately
described up to $U/J\approx 3\dots 4$, we will likewise analyze the stopping dynamics
in Sec.~\ref{sec.stopping.dyn} up to this regime of interaction strengths.


\section{\label{sec.stopping.dyn}Stopping dynamics}

\subsection{\label{subsec.energy.loss}Energy loss of the projectile}

We now simulate collisions of protons ($Z_\textup{p}=1$) with honeycomb clusters of size
$L=24$ and $54$. To characterize the stopping dynamics, we consider different impact kinetic
energies $E_\textup{kin}=\tfrac{1}{2}m_\textup{p}\dot{\vec{r}}_\textup{p}^2(t=0)$, ranging
from below $1$\,keV to about $0.5$\,MeV, and measure the energy loss $S_\textup{e}$, defined
as the change of the projectile's kinetic energy after passing through the lattice:
\begin{align}
\label{eq.energyloss}
 S_\textup{e}=E_\textup{kin}(t=0)-E_\textup{kin}(t\rightarrow\infty)\,.
\end{align}
This quantity is proportional to the commonly used stopping power (dissipated power per
length). We specify the kinetic energy of the proton in units of keV/u, where u denotes
the unified atomic mass unit.

\begin{figure}[hptb]
\includegraphics[width=0.483\textwidth]{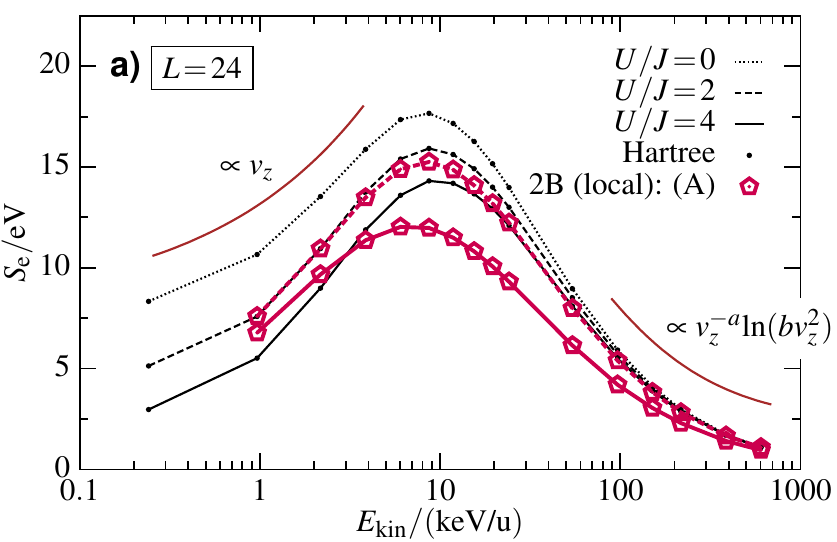}
\includegraphics[width=0.483\textwidth]{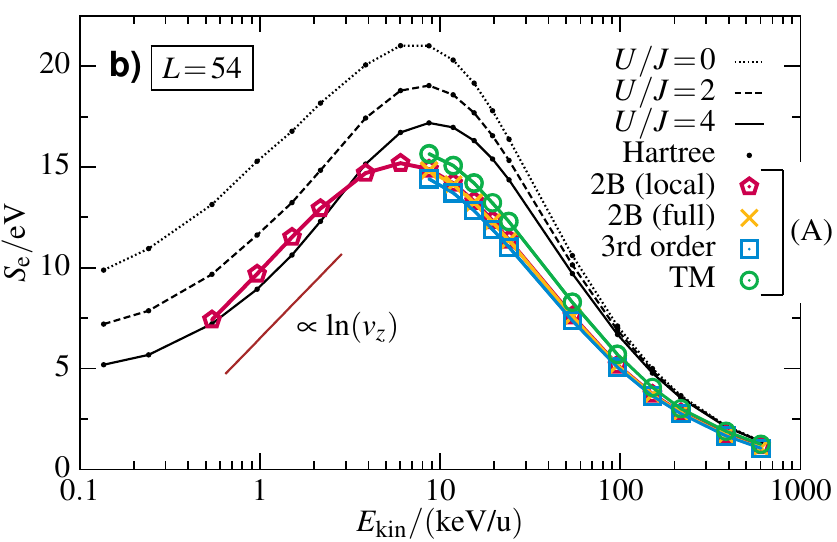}
\includegraphics[width=0.483\textwidth]{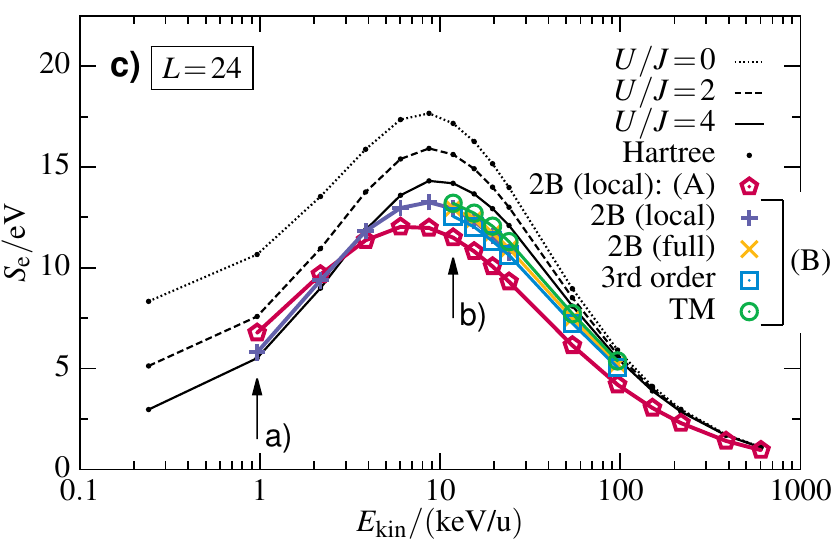}
 \caption{(Color online) Energy loss $S_\textup{e}$ for protons passing through honeycomb
 clusters of size $L=24$ (panels~a) and~c)) and $L=54$ (panel~b)). In all panels, the value
 of the on-site interaction $U/J$ is encoded in the line style, and the black lines indicate
 the results of the Hartree approximation. In panel~a), the red curves show the energy loss
 in the local second Born approximation with initial state~(A). In panel~b), we present the
 same analysis for the larger cluster, including results of the full second Born (yellow),
 the third-order (blue) and the T-matrix approximation (green). Panel~c) shows the influence
 of initial correlations, comparing the second Born results of panel~a) to local and full
 second Born, third-order as well as T-matrix calculations for the initial state~(B). The
 arrows in panel~c) indicate the two situations analyzed in more detail in Fig.~\ref{fig.stopping.explanation}.}
 \label{fig.stopping.lattice}
\end{figure}

We begin with the analysis of $S_\textup{e}$ for the smaller cluster, see Fig.~\ref{fig.stopping.lattice}a.
In Hartree approximation (black lines), we find a pronounced maximum of the energy loss
in the regime of considered energies, which is the behavior known from the stopping power
of nonrelativistic ions (we note, however, that the position of the maximum is typically at
larger energies, compare with Sec.~\ref{sec.graphene}). When $U$ is increased, the
peak height decreases and, at the same time, the peak slightly shifts towards larger
proton energies. At large impact energies, the curves for different interaction strength
approach each other, leading to a rather generic scaling of the energy loss as
$\propto v_z^{-a}\textup{ln}(b v_z^2)$ (with fit parameters $a,b>0$). Thus, the
high-energy tail is consistent with predictions from the non-relativistic Bethe
formula~\cite{sigmund06}. On the other hand, for low energies, the  change of the
energy loss is closer to $\sim v_z$.

Next, we examine the influence of electron-electron correlations, cf.~the red curves in
Fig.~\ref{fig.stopping.lattice}a, which represent local second Born calculations for
the initial state (A) of Sec.~\ref{subsec.mb.approx}, where the lattice system has
equilibrated before the impact of the proton. For small interactions $U/J\lesssim2$, we
find that corrections to the Hartree approximation are rather small. On the contrary, for
$U/J=4$, we observe clear deviations from the mean-field picture, with a decrease of
$S_\textup{e}$ over a large energy window and a slight increase around $E_\textup{kin}\approx1$\,keV/u.

In Fig.~\ref{fig.stopping.lattice}b, we present the same analysis for the larger honeycomb
cluster with $L=54$ sites, including results for various approximations of the self-energy.
While $S_\textup{e}$ becomes generally larger compared to the smaller cluster, we notice
that correlations have the same effect of reducing the energy loss for proton energies of
$E_\textup{kin}\gtrsim5$\,keV/u, as was observed for $L=24$. At the same time, the low-energy
tail behaves differently: here we find a scaling $\propto\textup{ln}(v_z)$). Moreover,
we observe that all considered self-energies lead to very similar stopping results. In
particular, there is very good agreement between the local and full second-order Born
approximation, which indicates that here it is sufficient to treat correlations locally.
We emphasize again that a non-local self-energy (with $\Sigma_{ij}\neq0$) or a more complex
self-energy (including higher-order diagrams) generally brings about a drastic increase of
the computation time, particularly on a large time grid, which is required to study the
impact of slow projectiles. For this reason, we show results beyond the local second Born
approximation in Fig.~\ref{fig.stopping.lattice}b only for correspondingly large proton energies.

Standard stopping power calculations of a charged particle usually consider the target
material in the ground state before the collision. This is, however, not the case for our
simulations with the initial condition (A) and self-energies beyond mean field. To quantify
the effect of this systematic inconsistency, we repeat some of the simulations with the initial
state~(B), see Fig.~\ref{fig.stopping.lattice}c for $L=24$ and $U/J=4$. As a result, we
observe that the form of the correlated initial state has a non-negligible influence on
the energy loss of the projectile. In fact, we find that the self-consistent correlated
ground state [case (B)] yields energy losses which are overall closer to those of the Hartree
approximation. Nevertheless, there remain significant differences between correlated and 
mean-field calculations, most importantly around the maximum of the curves.

\subsection{\label{subsec.td.density}Time-dependent density response of the electron system}

\begin{figure}[t]
\includegraphics[width=0.483\textwidth]{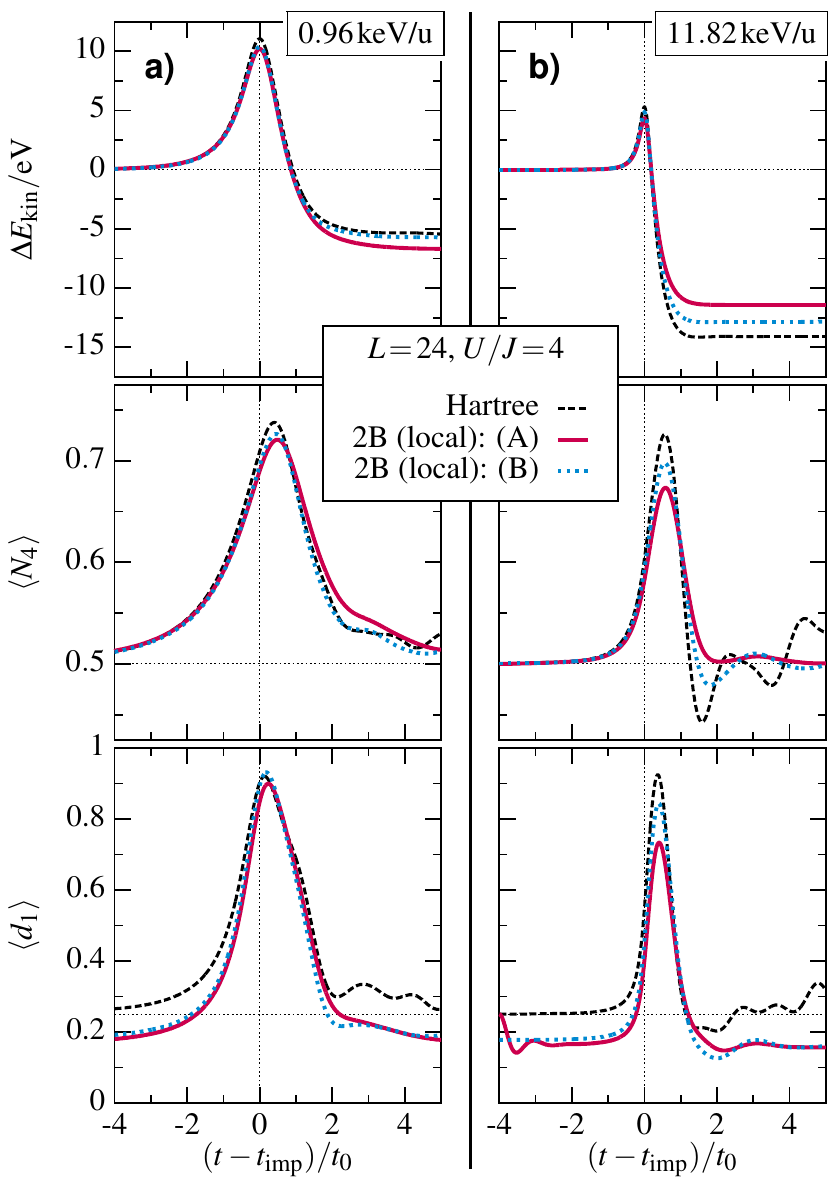}
 \caption{(Color online) Coupled proton-electron dynamics for the honeycomb cluster with $L=24$
 sites for two different proton energies, $E_\textup{kin}=0.96$\,keV/u (left panels) and $11.82$\,keV/u
 (right panels), corresponding to the arrows shown in Fig.~\ref{fig.stopping.lattice}c. The coupling
 strength is fixed, $U/J=4$, and three many-body approximations are compared (see inset). Top panels
 change of the projectile's kinetic energy $\Delta E_\textup{kin}(t)=E_\textup{kin}(t)-E_\textup{kin}(t=0)$
 as function of time. Center panels: time evolution of the electron density averaged over the central
 sites labeled $1$ to $4$ in Fig.~\ref{fig.lattice}, i.e., $\langle N_4\rangle(t)=\tfrac{1}{4}\sum_{i=1}^{4}\langle n_{i\sigma}\rangle(t)$.
 Bottom panels: time evolution of the double occupation $\langle d_1\rangle(t)=\langle n_{1\uparrow}n_{1\downarrow}\rangle(t)$
 on the site $1$, which is closest to the impact point of the projectile.}
 \label{fig.stopping.explanation}
\end{figure}

To gain insight into the effect of correlations on $S_\textup{e}$ and the physical mechanisms,
we now analyze the response of the lattice electrons to the approaching projectile for a fixed
value of the interaction strength, $U/J=4$. The general scenario is as follows. During the
early stage of the dynamics, the electrons (initially distributed uniformly over the cluster
with $\langle n_{i\sigma}\rangle=0.5$) start to accumulate close to the impact point and, thus,
create a negative net space charge, which attracts and accelerates the proton towards the cluster.
After passing through the lattice plane, the proton then looses kinetic energy, depending on the
non-adiabatic response of the electron density. For two different proton energies [indicated by
arrows labeled a) and b) in Fig.~\ref{fig.stopping.lattice}c] the precise dynamics is shown in
Fig.~\ref{fig.stopping.explanation}. There, we compare the Hartree approximation to the local
second Born approximation for both considered initial states (A) and (B). 

The difference in the time scale on which the observables change during the collision process
is evident:~While at a kinetic energy around $1$\,keV/u, the electron density and the double
occupation in the center of the honeycomb cluster change on a time scale of a few inverse
hopping times, $t_0=\hbar/J$, they change on a time scale comparable to $t_0$ for the much
faster proton ($\sim10$\,keV/u). This difference has immediate consequences for the energy
transfer to the lattice: From Fig.~\ref{fig.stopping.explanation}b (fast proton), we find
that the exchange of energy between projectile and target occurs mainly during the stage of
electron accumulation. Together with a retarded response of the electron density in the second
Born approximation [dotted and solid lines in panel b)], this translates into a faster proton
(of a few eV) after the collision, as compared to the mean-field calculation (dashed line).
On the contrary, for the slow proton, the energy loss is defined by both the buildup \textit{and}
the removal stage of the charge-induced confinement potential. For this reason it, is not a
priori obvious how $S_\textup{e}$ is altered by correlations. This is also confirmed by the
difference of the two second Born calculations, cf.~in particular the center panel in Fig.~\ref{fig.stopping.explanation}a.
Here, the calculation which starts from the correlated ground state (dotted line) shows a
density response rather close to the Hartree approximation~\cite{densitymatrix}, whereas
the simulation which uses the equilibrated Hartree ground state as initial state (solid line)
yields an energy transfer that is clearly larger than the mean-field result. The time evolution
of the double occupation is, however, almost identical in both correlated cases, but significantly
different from the mean-field approximation, see the bottom panel in Fig.~\ref{fig.stopping.explanation}a.

\subsection{\label{subsec.td.spectrum}Time-dependent electron spectral properties}

\begin{figure}[t]
\includegraphics[width=0.483\textwidth]{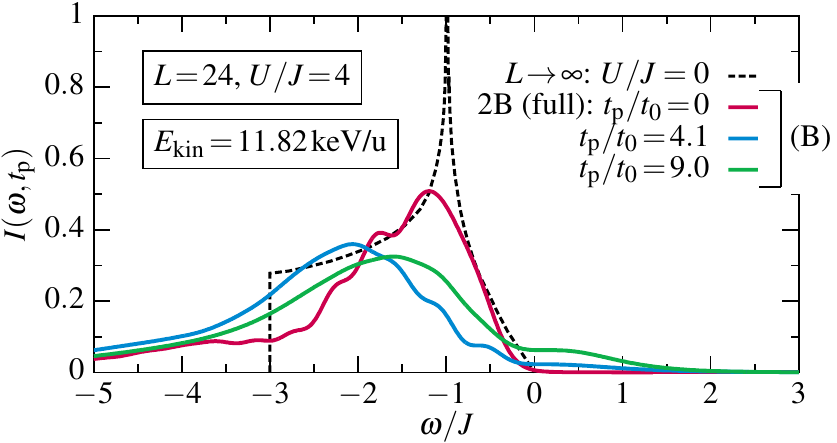}
 \caption{(Color online) Time-resolved photoemission spectrum $I(\omega,t_\textup{p})={I_0^{-1}}\sum_{i=1}^{L}I_i(\omega,t_\textup{p})$
 of a cluster with $L=24$ and $U/J=4$ for the stopping dynamics of a proton with energy $E_\textup{kin}=11.82\,$keV/u
 (the scenario is similar to Fig.~\ref{fig.stopping.explanation}b). The calculations are performed
 in full second Born approximation with a correlated initial state [approximation (B)]. Red curve:
 initial state, black dashes: ground state of an infinite system at $U/J=0$. Blue line: spectrum
 when the projectile passes through the lattice plane. Green line: ``final'' state after the collision.
 To resolve the spectrum at the different stages of the dynamics, we have set the probe pulse width
 here to $\tau=2t_0$, cf.~Appendix~\ref{sec.appendix.spectrum}.}
 \label{fig.stopping.explanation.pes}
\end{figure}

An even closer look at the electronic excitations during the collision process is provided
by the time-resolved photoemission spectrum $I(\omega,t_\textup{p})=I_0^{-1}\sum_{i=1}^{L}I_i(\omega,t_\textup{p})$,
with normalization factor $I_0$. Our NEGF approach directly yields this quantity (see Appendix),
and we present the results in Fig.~\ref{fig.stopping.explanation.pes} for two different probe
times $t_\textup{p}$. Prior to the impact of the proton ($t_\textup{p}=0$), the spectrum
corresponds to the correlated ground state [we use initial state~(B)] of the system and is,
thus, analogous to the one discussed in Fig.~\ref{fig.equil.pes1}b. Note, however, that here
we use $U/J=4$, and we average over the whole cluster and use a different probe pulse. At a
later time, when the projectile just passes the lattice plane ($t_\textup{p}\sim4.1t_0$), we
observe a spectrum which indicates a strong redistribution of electrons in the lower Hubbard
band, particularly towards lower energies (blue line in Fig.~\ref{fig.stopping.explanation.pes}).
This redistribution is, obviously, a result of the negative electronic confinement potential
induced to the lattice electrons by the projectile and corresponds to a net energy loss of
the electron system. Finally, at time $t_\textup{p}\gtrsim9t_0$, the proton has passed through
the lattice and is located far enough such that it does not affect the electrons anymore. We,
therefore, measure a state of the electrons that is close to the ``final'' state. This state
is characterized by a net energy gain of the electron system (as was shown in the top panel
of Fig.~\ref{fig.stopping.explanation}b). Here, we can resolve the spectral distribution
of this energy: a substantial amount of electrons is being excited (above the Fermi level,
$\omega_\textup{F}=0$) into the upper Hubbard band. Of course, on a longer time scale,
(part of) this energy will be transferred from the electrons to lattice vibrations (phonons),
but this is beyond the present model.

\subsection{\label{subsec.gkba}Projectile energy loss within the Generalized Kadanoff--Baym ansatz}

\begin{figure}[b]
\includegraphics[width=0.483\textwidth]{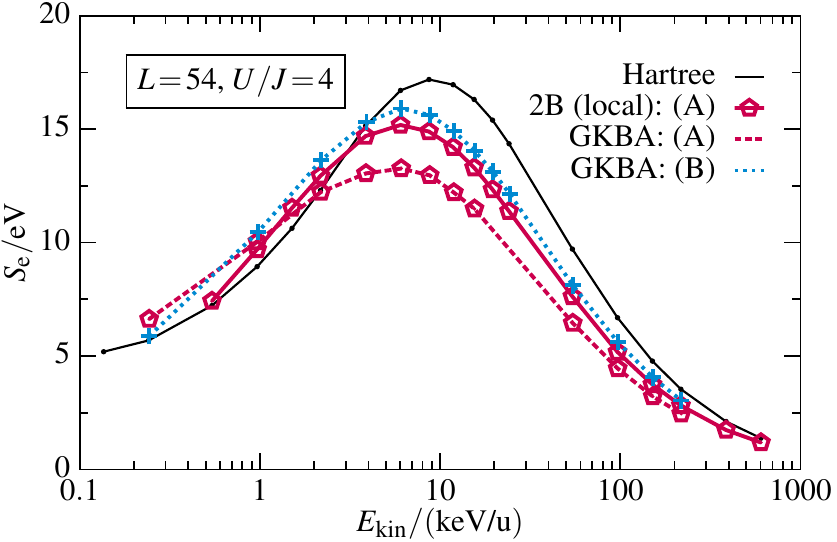}
 \caption{(Color online) Energy loss for the honeycomb cluster with $L=54$ sites in second
 Born approximation. Comparison of the full two-time simulation and the GKBA, see text for
 details. Black and red solid lines are the same as in Fig.~\ref{fig.stopping.lattice}b;
 red dashed and blue dotted curves: GKBA results for the initial states (A) and (B), respectively.}
 \label{fig.stopping.gkba}
\end{figure}

In this section, we analyze the generalized Kadanoff--Baym ansatz (GKBA) that was discussed
in Sec.~\ref{subsec.mb.approx}. This approximation has recently attracted growing
attention~\cite{latini14,hermanns14,perfetto15,schluenzen16.cpp}, because it provides a way
to significantly reduce the numerical effort of the computation of the NEGF, while still
preserving the conservation laws of the chosen many-body approximations. Here, we apply the
GKBA to reconstruct the two-time Green function $G_{ij\sigma}(t,t')$ for  second Born self-energies
from its time-diagonal value by using Hartree propagators, see Ref.~\cite{schluenzen16.cpp}
for details. For the present setup, the GKBA allows us to extend the full second Born calculations
of Fig.~\ref{fig.stopping.lattice}b towards significantly longer times and, thus, to lower proton
impact energies. We also note that, for finite systems which are strongly excited, the GKBA has
been found to be free of certain artifacts of the two-time simulations \cite{hermanns14}, while
being of comparable accuracy than the latter. 

In Fig.~\ref{fig.stopping.gkba}, we present such GKBA simulations for both initial states, (A)
and (B), as discussed above. For the initial state (A), we find a qualitative agreement with
the analogous two-time calculations. At the same time, the GKBA simulations yield a systematically
lower energy loss $S_\textup{e}$ for proton energies of $E_\textup{kin}\gtrsim5$\,keV/u than
the two-time simulations. In the present case, the projectile induces a rather strong and non-local
perturbation, which is typically well described by the GKBA \cite{hermanns14}. Whether the GKBA
or two-time results for the stopping power are more accurate is presently unknown, as there are
no exact results available, and this remains to be resolved in future studies. 

Finally, we perform GKBA simulations with the fully correlated initial state (B). This leads to
significantly increased results for the  energy loss spectrum of the protons (blue dotted curve),
which are closer to the mean-field result. The most striking achievement is that the GKBA
simulations can be extended towards projectile energies around $200$\,eV. Interestingly, for
these energies, the stopping power is significantly increased, as compared to the mean-field
result. At the same time, with the use of Hartree propagators, we loose direct access to the
correlated spectral functions.

\subsection{\label{subsec.finite.temp}Finite temperatures}

For slow projectiles, we have seen in Secs.~\ref{subsec.energy.loss} and~\ref{subsec.gkba},
that the inclusion of electron-electron correlations can lead to a slight increase of the energy
loss in comparison to the mean-field treatment of the collision process. As this effect seems
to be larger for a lattice system which is initially not in the self-consistent ground state [and
thus has a non-zero effective temperature, cf. initial state (A)], it is worthwhile to discuss
in more detail the influence of a finite electron temperature on the stopping dynamics.

\begin{figure}[t]
\includegraphics[width=0.483\textwidth]{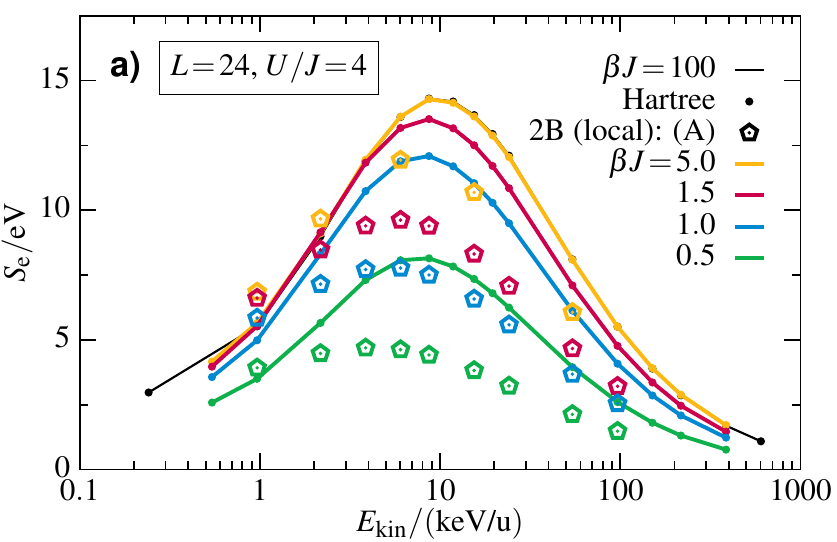}
\includegraphics[width=0.483\textwidth]{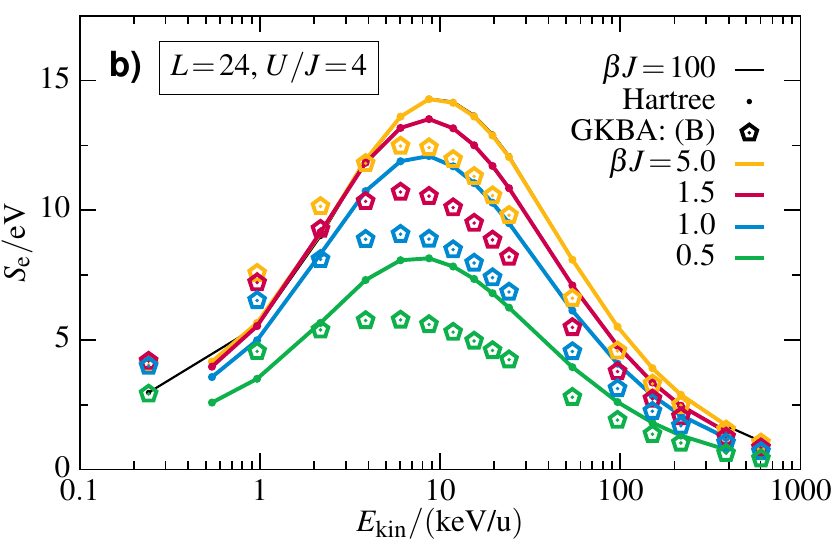}
 \caption{(Color online) Temperature dependence of the energy loss $S_\textup{e}$ for the honeycomb
 cluster with $L=24$ sites and $U/J=4$. Black lines: zero-temperature Hartree calculations, as in
 Fig.~\ref{fig.stopping.lattice}a. Colored lines: Hartree results for different temperatures.
 Symbols: local second Born results for the same temperatures $\beta J=5$, $1.5$, $1$ and $0.5$.
 Panel a): Local second Born calculations with initial state~(A). Panel b): GKBA calculations as
 in Sec.~\ref{subsec.gkba} with initial state~(B).}
 \label{fig.stopping.temperature}
\end{figure}

\begin{figure}[t]
\includegraphics[width=0.483\textwidth]{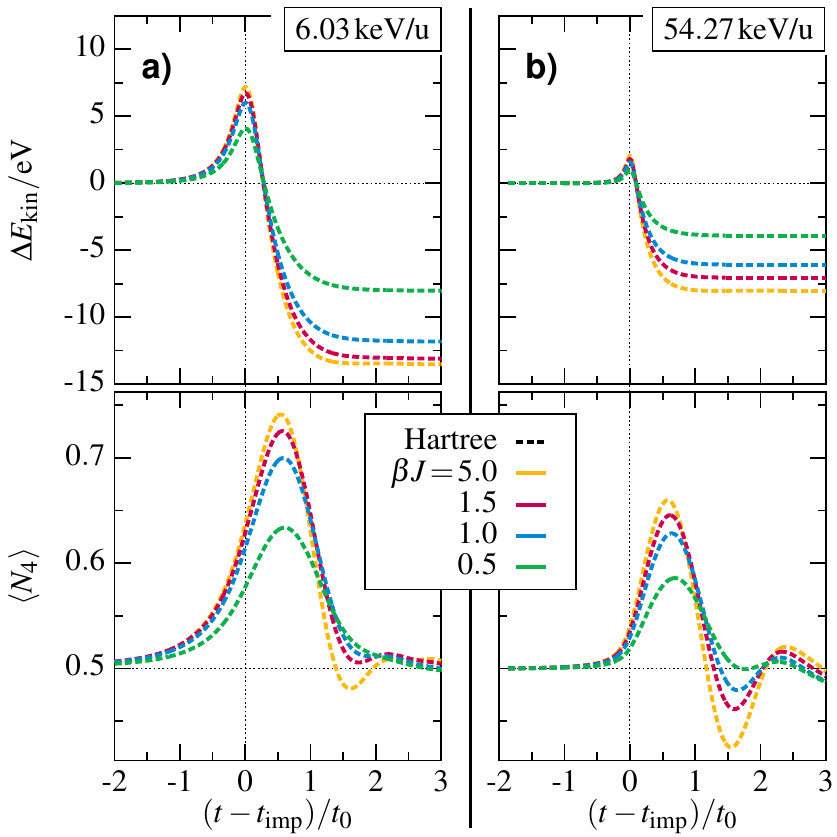}
 \caption{(Color online) Time-dependent energy change $\Delta E_\textup{kin}(t)=E_\textup{kin}(t)-E_\textup{kin}(t=0)$
 of the proton (upper panel) and electron density $\langle N_4\rangle(t)=\tfrac{1}{4}\sum_{i=1}^{4}\langle n_{i\sigma}\rangle(t)$
 at the four sites around the impact point (lower panels) for four different temperatures. Same
 cases (Hartree dynamics, $U/J=4$) as shown in Fig.~\ref{fig.stopping.temperature}.}
 \label{fig.stopping.temperature1}
\end{figure}

In contrast to other approaches, the effect of finite temperature is straightforwardly
incorporated in the NEGF formalism, where temperature effects enter the KBE via the
initial state defined in Eq.~(\ref{eq.hartree}). In Fig.~\ref{fig.stopping.temperature},
we perform Hartree and second Born calculations for $L=24$ and different inverse temperatures
$\beta J\ll100$.  For $\beta J=5$, which corresponds to an electron temperature of
$k_\textup{B}T=0.56$\,eV (or about $T=6500$\,K) for a hopping amplitude of $J=2.8$\,eV,
we measure energy loss spectra (cf.~the yellow curves) that are still very close to the
ground state results of Fig.~\ref{fig.stopping.lattice}a. For higher temperatures, $\beta J<5$,
on the other hand, we observe that the energy loss systematically decreases with temperature,
whereas the maximum of the spectrum shifts to slightly lower energies. These trends continue
even for higher electron temperatures (obviously, this refers to a nonequilibrium state,
where the electron temperature is decoupled from the lattice), see the red, blue and green
solid curves. 

To understand the origin of the reduction of the energy loss with temperature, we investigate
in Fig.~\ref{fig.stopping.temperature1} the time dependence of the proton energy and of the
local electron density computed in mean-field approximation. Obviously, a temperature increase
reduces the local enhancement of the electron density, as thermal fluctuations reduce the
coherent response of the electrons to the projectile. 

It is interesting to note that electron correlations are important even at the highest
temperature considered [$\beta J=0.5$ ($T\approx65000$\,K for $J=2.8$\,eV)]. Here, compared
to the mean-field model, the local second Born calculations yield a clear shift of the whole
spectrum towards smaller energies. Furthermore, the maximum energy loss is substantially
reduced. Also, with reduction of the temperature, the relative importance of correlations
seems to increase. At the highest temperature, the peak value of the second Born calculation
is only about half of the mean-field result, cf.~Fig.~\ref{fig.stopping.temperature}a.
This unexpected behavior is due the reduction of quantum diffraction effects with temperature,
leading to an increased electron localization, which will be investigated in more detail
elsewhere.


\section{\label{sec.graphene}Application to graphene}

As a supplementary investigation, we examine, in this section, whether the coupled
NEGF--Ehrenfest approach can be applied also to study the collision of charged particles
with real (low-dimensional) materials. As example, we consider a two-dimensional sheet
of graphene. As was shown in Refs.~\cite{wehling11,tang15}, the equilibrium properties
of graphene~\cite{neto09,katsnelson12} are well described through an extended Hubbard
model with a nearest-neighbor hopping $J$ on the honeycomb lattice using, beyond the
on-site interaction $U$, additional non-local Coulomb interactions $V_{ij}$, that are
known to stabilize the Dirac semimetallic phase~\cite{wu14}. However, it is not clear
a priori whether this model holds also out of equilibrium. In particular, the present
situation of the impact of a charged particle corresponds to a (locally) very strong
excitation, driving the system far away from equilibrium. This question can only be
answered by direct simulations of this process and by comparison to reliable reference
data for the stopping power.  

In order to map this extended Hubbard model to the Hamiltonian of the form~(\ref{eq.ham1})
with purely local interactions, we follow Ref.~\cite{schueler13} and use an effective on-site
interaction $U'=U-\bar{V}=1.6J$, where $\bar{V}$ denotes a weighted average over the
non-local contributions. Although this approximation has limitations, e.g.,~\cite{tang15},
it is agreed to be, at least, qualitatively correct. Moreover, we extend the Hamiltonian
of Sec.~\ref{sec.comp.setup} in two regards:
\begin{enumerate}
\item[(i)] We take into account the existence of four valence electrons per site. This
means we consider (instead of a single-band model) a system with four independent Hubbard
bands of equal hopping and interaction parameters, which together describe the dynamics
of the four electrons provided by each $sp^2$-hybridized carbon atom in the graphene sheet.
We are aware of the fact that such an approach excludes the specific nature of the $\sigma$-
and $\pi$-bonds as well as possible \mbox{($sp$-)}interband transitions. The main advantage
of this model is however, that it can be straightforwardly implemented by setting the local
net charges $Z_i$ in Eq.~(\ref{eq.ham3}) to $Z_i=4(1-\sum_\sigma \langle n_{i\sigma}\rangle)$,
leaving open a single parameter, the hopping amplitude $J$, which we will use below to
adjust the maximum energy transfer.
\item[(ii)] We account for the fact that the incident projectile can influence the electron
mobility on the lattice. This includes local changes to the electron's kinetic energy which
originate from the presence of the off-diagonal matrix elements of the interaction potential
$W_{ij}$ between the projectile and the electrons on the lattice. Below, we approximate such
a renormalization of the hopping to be proportional to the average potential energy between
neighboring sites, i.e., we define an effective time- and site-dependent hopping amplitude
\begin{align}
 \label{eq.hoppingrenormalization}
 J_{ij}(t)&=\left\{\begin{array}{cc}
-J+W_{ij}(t) &\textup{,\;} |i-j|=1\\[0.5pc]
0&\textup{,\;} \textup{otherwise}
\end{array}\right.\,,
\end{align}
where
\begin{align}
 W_{ij}(t)&=\gamma \frac{W_{ii}(t)+W_{jj}(t)}{2}\,.
\end{align}
The proportionality factor $\gamma$ can be interpreted as the strength of the orbital
overlap and will be used as a second fit parameter below (see Appendix~\ref{sec.appendix.graphene}
for details). 
\end{enumerate}

We note that the ansatz (\ref{eq.hoppingrenormalization}) neglects corrections of the
form $W_{ij}$ with $|i-j|>1$, which is justified because the wave functions of next-nearest
and more distant neighbors have in general a much smaller overlap. Nevertheless, a further
improved treatment of the off-diagonal components may be important for future studies,
since the projectile induces strong perturbations to the system.

\begin{figure}[t]
 \includegraphics[width=0.483\textwidth]{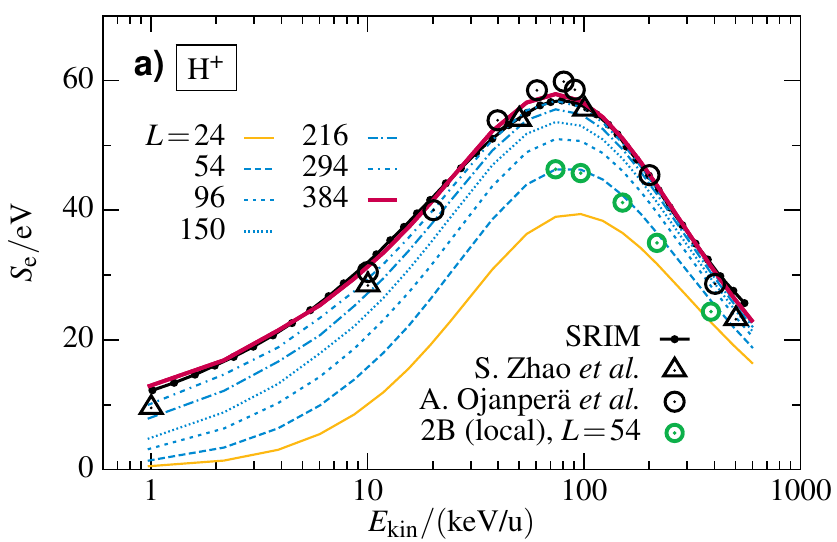}
 \includegraphics[width=0.483\textwidth]{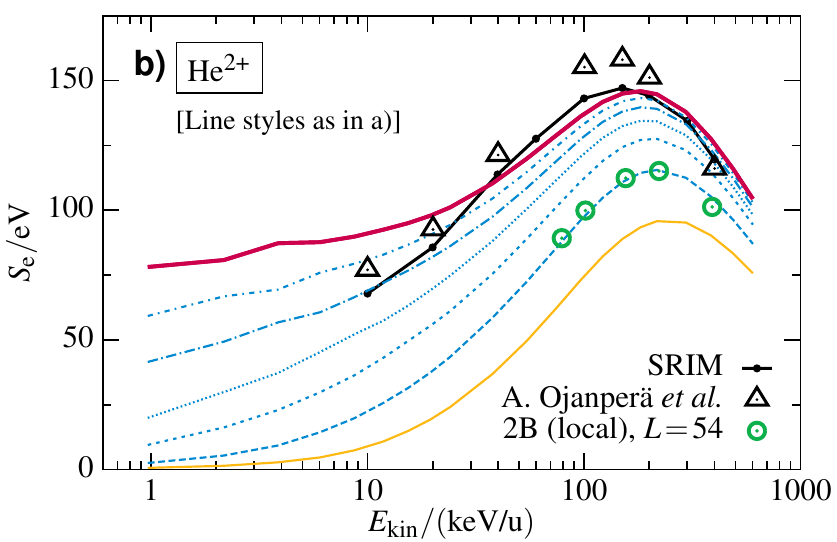}
 \caption{(Color online) Energy loss of a)~hydrogen ions (H$^+$) and b)~alpha particles
 (He$^{\textup{2+}}$) penetrating through a single layer of graphene. In both panels, the
 strength of the Coulomb interaction is $U/J=1.6$, the fit parameters are $J=3.15\,$eV and
 $\gamma=0.55$, and the initial $z$-position of the projectile is $z=20a_0$. The Hartree
 results for different cluster sizes, ranging from $L=24$ (bottom) to $L=384$ (top), are
 shown by different line styles. For the cluster with $L=54$ sites, we also performed second
 Born calculations (green circles) showing that correlation corrections are rather small
 in this case. For a detailed discussion of the $U$-, $J$-, and $\gamma$-dependence, see
 Appendix~\ref{sec.appendix.graphene}. The black symbols and lines correspond
 to ab-initio TDDFT calculations and SRIM data, respectively (taken from Refs.~\cite{zhao15,ojanperae14}).}
\label{fig.stopping.graphene.revision}
\end{figure}

In Fig.~\ref{fig.stopping.graphene.revision}, we present stopping results for protons
and alpha particles ($Z_\textup{p}=2$) for the model parameters \mbox{$J=3.15\,$eV} and \mbox{$\gamma=0.55$}.
To obtain reasonable agreement with ab-initio TDDFT and SRIM data for the planar
{\em infinitely extended} graphene sheet~\cite{ojanperae14,zhao15}, we consider cluster
sizes as large as $L=384$, which are easily treated in Hartree approximation. The neglect
of correlations is justified due to the relatively small on-site interaction strength of
$U/J=1.6$, recall Sec.~\ref{subsec.energy.loss}. This expectation is confirmed by
performing additional local second Born simulations for $L=54$ (green circles) that lie
on top of the Hartree curves, cf. Fig.~\ref{fig.stopping.graphene.revision}a and \ref{fig.stopping.graphene.revision}b.
Generally, we find that the energy transfer increases with the cluster size, which is
consistent with results for graphene clusters discussed in Ref.~\cite{bubin12}. In
Fig.~\ref{fig.stopping.graphene.revision}a, the curves $S_\textup{e}(E_\textup{kin})$
tend to converge for large $L$ and, for practically all considered proton energies,
well approach the energy loss given by the reference data. 

Finally, an important test of the model is provided by Fig.~\ref{fig.stopping.graphene.revision}b.
There, we use exactly the same model parameters  $J$ and $\gamma$ to simulate the energy
loss for collisions of single-layer graphene with bare helium nuclei (He$^{2+}$). Without
further adjustments, we recover good agreement with the available reference data, including
the increase of the overall magnitude of $S_\textup{e}$ compared to the case of protons
and, in addition, the shift of the maximum energy loss towards larger kinetic energies.

Given the simplicity of the model Hamiltonian, it is interesting, that our NEGF-based
approach reveals the correct trends for a fairly realistic system. On the other hand,
however, we have to note also problems of the model. In particular, at the low-energy
tail of the energy loss curve, we observe significantly larger values compared to the
TDDFT simulation in the case of alpha particles, cf.~the red curve in Fig.~\ref{fig.stopping.graphene.revision}b. 
The origin of these discrepancies are not fully clear yet, and, therefore, in this range,
additional correlated simulations as well as improvements to the model are required in
the future.


\section{\label{sec.conclusions}Conclusions}

In summary, we have presented a combined nonequilibrium Green functions and classical
Ehrenfest dynamics approach to the interaction of a nonrelativistic charged particle
with a (strongly) correlated system. Our approach allows for a fully time-dependent
treatment and is, thus, able to resolve non-adiabatic processes in the electronic sub-systems.
To explore the role of electronic correlations, we performed solutions of the two-time
Keldysh--Kadanoff--Baym equations using different many-body approximations for the self-energy:
the second Born, third-order and the T-matrix approximations. This enabled us to demonstrate
that electron-electron correlations do significantly influence the slowing down of a charged
projectile in, both, the low and high-energy limits. The high computational effort of the
NEGF simulations has limited us to projectile energies of $1$\,keV, as lower impact energies
increase the interaction time with the lattice and, in turn, the computing time. To extend
the simulations to lower energies, we have applied the generalized Kadanoff--Baym ansatz
(GKBA), which is substantially more efficient. Interestingly, these simulations predict
an energy loss well above the mean field model indicating that correlations can enhance
the slowing down of a (slow) projectile. How accurate these results are is not known at the moment.
This requires further analysis via full two-time simulations, the use of improved self-energies
such as T-matrix self-energies, as well as independent TDDFT simulations.

Of particular current interest is the energy loss of low-energy (below $1$\,keV) charged
particles in solid materials. An important field of applications are low-temperature plasmas.
Questions of interest include the stopping power in materials with very strong electronic
correlations (e.g., lattice models with $U/J \gtrsim 10$) or for magnetically ordered systems
or insulators, where the stopping power can vanish below a certain threshold~\cite{markin09}.
Furthermore, it will be important to extend the model beyond the Hubbard model to better
capture realistic material properties, e.g., by using a Kohn-Sham basis. This, however,
will drastically increase the computational requirements.

Additional questions of interest at low energies concern the inclusion of all relevant
dissipation mechanisms, in particular, inelastic mechanisms such as phonons, impact
excitation and ionization or re-emission of particles. Further relevant processes
include neutralization of the ion before impact and capture (sticking), which is expected
to cause deviations from the linear velocity scaling. Finally, it will be important to
also consider more complex charged projectiles that are different from bare ionic cores.
Here, intra-ionic electronic excitations play an important role in the stopping dynamics,
e.g.,~\cite{pamperin15}. 

From a technological point of view, it would furthermore be interesting to explore whether
the energy deposition can be externally controlled, e.g., by time-dependent (laser) fields,
which excite the target material before or during the impact. The potential effect of such an
out-of-equilibrium situation was demonstrated by an analysis of an increased temperature of
the electronic sub-system. For such kinds of nonequilibrium investigations, our NEGF-based
approach represents an optimal toolbox, as it handles external fields self-consistently and
non-perturbatively and can include arbitrary scattering processes in a systematic way.


\acknowledgments

We thank Andrea Marini and Davide Sangalli for stimulating discussions and Lasse Wulff for
performing simulations during the early stage of this work.


\appendix

\section{\label{sec.appendix.spectrum}Time-resolved photoemission spectrum}

As an essential test for the numerics and the time propagation of the KBE~(\ref{eq.kbe}),
we verify here, whether the correct spectrum and bandwidth are recovered from the two-time
NEGF in the limit of an infinite honeycomb lattice. To this end, we compute the photoemission
spectrum for vanishing on-site interaction [$U=0$ in Hamiltonian~(\ref{eq.ham1})] and consider
clusters of different size $L$.

\begin{figure}[b]
\includegraphics[width=0.483\textwidth]{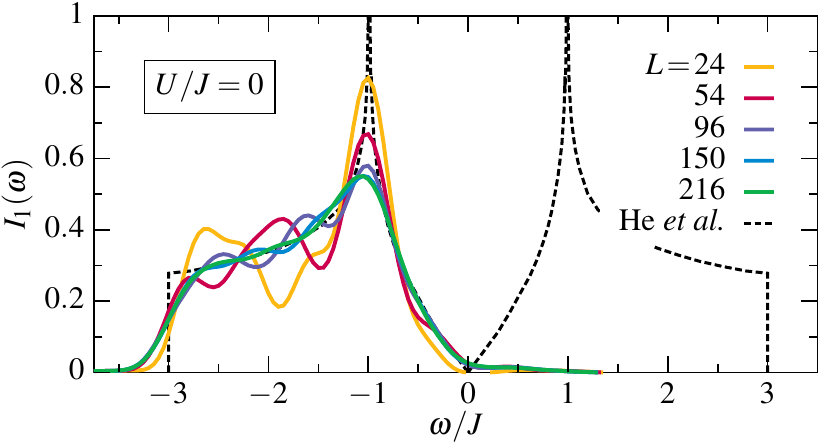}
 \caption{(Color online) Photoemission spectrum $I_1(\omega)$ for different uncorrelated
 honeycomb clusters of size $L$, measured at a probe time $t_\textup{p}=8t_0$ with $\tau/t_0=4$,
 cf.~Eq.~(\ref{eq.appendix.pes}). The black solid line shows the density of states in
 the valence and conduction band of the extended honeycomb lattice at half-filling, e.g.,
 Refs.~\cite{neto09,he12}.}
 \label{fig.appendix.pes1}
\end{figure}

We consider the photoemission signal of a reference site $i$, which is
given by~\cite{eckstein08.pes}
\begin{align}
\label{eq.appendix.pes}
 &I_i(\omega,t_\textup{p})\\
 &=-\ii\int\!\d t\!\int\!\d t'\, s(t-t_\textup{p})s(t'-t_\textup{p})\textup{e}^{\ii\omega(t-t')}G^<_{ii\sigma}(t,t')\nonumber
\end{align}
at some probe time $t_\textup{p}$. The function $s(t)$ thereby describes the envelope of
the probe pulse and is chosen to be of Gaussian form, i.e.,
\begin{align}
s(t)=\frac{1}{\tau\sqrt {2\pi}}e^{-t^2/(2\tau^2)}\,,
\end{align}
where we set $\tau=4 t_0$ with $t_0=\hbar/J$ and $J=2.8$\,eV.

Figure~\ref{fig.appendix.pes1} shows the spectrum $I_1(\omega)$ for honeycomb clusters with
up to $L=216$ sites at a probe time $t_\textup{p}=8t_0$, where, to evaluate the integral in
Eq.~(\ref{eq.appendix.pes}), we have computed the nonequilibrium Green function up to $t,t'=15 t_0$.
Aside from some pulse-induced peak broadening, we observe that the smaller clusters reveal
only a few single-particle states, which are due to the finite system size. On the other hand,
the spectra of the larger clusters ($L>96$ sites) approach already well the density of states
in the lower (valence) band of the extended honeycomb lattice, cf.~the black lines.

\section{\label{sec.appendix.graphene}Adaptation of the model to graphene}

\begin{figure}[t]
 \includegraphics[width=0.483\textwidth]{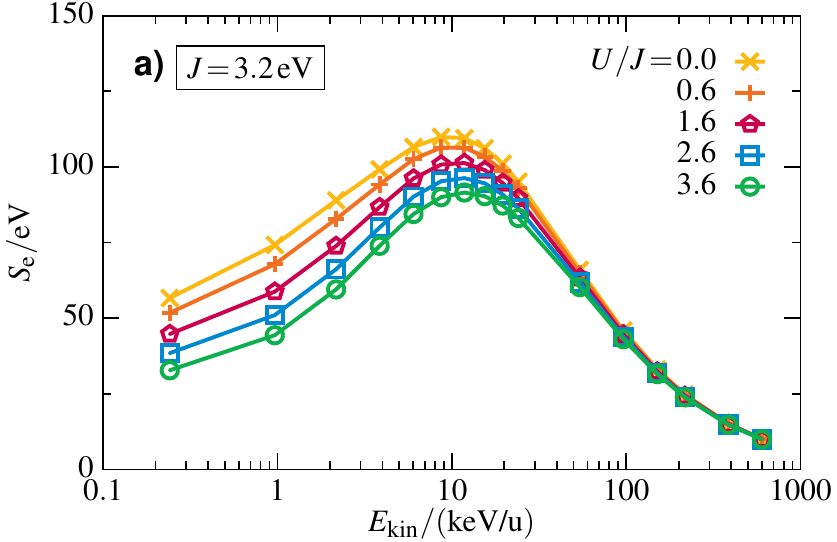}
 \includegraphics[width=0.483\textwidth]{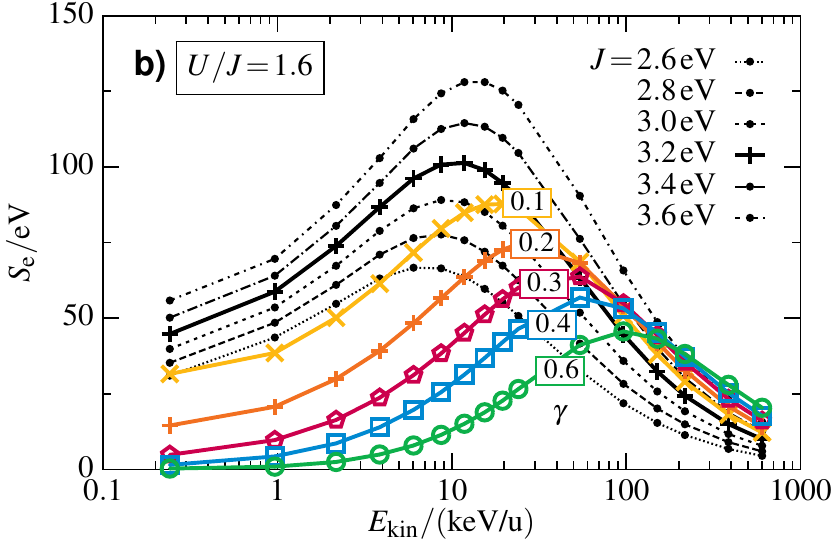}
 \caption{(Color online) Proton stopping dynamics as in Sec.~\ref{sec.graphene},
 but for different parameters $U$, $J$ and $\gamma$ in Hartree approximation. The system size
 is $L=54$. a)~Influence of the on-site interaction $U$ for fixed $J=3.2\,$eV and $\gamma=0$.
 b)~Influence of the hopping amplitude $J$ and the orbital overlap $\gamma$ for fixed
 interaction strength of $U/J=1.6$.}
 \label{fig.appendix.graphene}
\end{figure}

In order to model the impact of protons on a single sheet of graphene, we have tuned
in Sec.~\ref{sec.graphene} the hopping amplitude $J$ and the orbital overlap
parameter $\gamma$ [defined in Eq.~(\ref{eq.hoppingrenormalization})] such that the
energy loss spectrum is in good agreement with first principles and SRIM data. By
performing Hartree calculations, we show in Fig.~\ref{fig.appendix.graphene} in more
detail, how the energy transfer $S_e$ varies when these parameters are changed. Moreover,
we discuss the influence of the ratio of the on-site Coulomb interaction $U$ to the
hopping $J$.

From Fig.~\ref{fig.appendix.graphene}a, we observe that the interaction strength
mainly influences the energy transfer below about \mbox{$20\,$keV/u}, whereas the
high-energy tail remains unchanged. In the course of this, the low-energy tail as
well as the maximum energy loss increase fairly linearly with $U/J$. Fig.~\ref{fig.appendix.graphene}b
shows that the hopping amplitude and the overlap parameter $\gamma$ affect the energy
transfer substantially more than the Coulomb interaction. For a fixed value $U/J=1.6$,
a larger value of $J$ (i.e., an increase of the general electron mobility on the
lattice) leads to a larger energy loss, independently of the initial kinetic energy
of the projectile. On the other hand, if we increase the hopping locally by
choosing $\gamma>0$, we find that the energy transfer becomes considerably smaller,
which is accompanied by a shift of the maximum of $S_\textup{e}$ towards larger
kinetic energies.



\begin{thebibliography}{101}
\bibitem{phsymmetry} We note that in the case of an external excitation (as the field of a charged projectile) the
particle-hole symmetry breaks and the second Born self-energy becomes $\mathcal{O}\left(U^2\right)$. Nevertheless, the
Hubbard sites remain predominantly half-filled which makes the error in $\mathcal{O}\left(U^3\right)$ small.
\bibitem{densitymatrix} In this context, we note that the density matrix $\langle\rho_{ij\sigma}\rangle$ of the
correlated ground state (e.g., in the local second Born approximation [case~(B)]) is more similar to that of the
Hartree ground state than to that of the equilibrated Hartree ground state [case~(A)].
\end{thebibliography}
\end{document}